# The Place of Recurrent Novae among the Symbiotic Stars

J. Mikołajewska

*N. Copernicus Astronomical Center, Bartycka 18, 00-716 Warsaw, Poland*

**Abstract.** The observational properties of recurrent novae indicate that they can be divided into two subclasses:systems with a dwarf and a red giant secondary, respectively. The second type – which includes RS Oph – bears many similarities to symbiotic stars.

## 1. Introduction

Symbiotic stars are interacting binaries in which an evolved cool giant – either a normal M giant in S-types or a Mira variable embedded in an optically thick dust shell in D-types – transfers material to a hot white dwarf. In some cases the M giant is replaced by a G-K giant – these are called yellow symbiotic stars, and the white dwarf is replaced by a neutron stars – these are also classified as low mass X-ray binaries (e.g. V2116 Oph/GX 1+4). The presence of an evolved giant is essential to form a symbiotic binary, and so there must be enough space in the system to accomodate such a big star. Thus, symbiotic stars have the longest orbital periods and the largest component separations among the interacting binaries, and they are a very attractive laboratory for studying various aspects of interactions and stellar evolution in binary systems (e.g. Corradi, Mikołajewska & Mahoney 2003) .

The subclass of recurrent novae with giant secondaries – RS Oph, T CrB, V3890 Sgr and V745 Sco – have the same composition as the symbiotic binaries, and so they share many physical characteristics with these systems. The aims of this paper are to present the state-of-the-art in understanding of symbiotic binaries, and to discuss the place of these symbiotic recurrent novae (SyRNe) among these systems. In addition to the orbital and stellar parameters, the mechanisms of mass loss and accretion as well as the link between the SyRNe and the Z And-type systems will be discussed.

## 2. Orbital parameters

The distributions of the orbital periods, mass ratios and the stellar component masses for symbiotic stars have been recently discussed by Mikołajewska (2007). Since then, orbital period have been found for another 5 galactic systems, and spectroscopic orbits have been derived for 3 systems (Fekel et al. 2007; Gromadzki et al. 2007) which increased the number of systems with known orbital periods and spectroscopic orbits derived from the cool giant absorption features to 70 and 32 objects, respectively. The updated distributions are shown in Fig. 1, and they do not affect the previous conclusions. In particular, there is no systematic difference between the symbiotic systems in the galactic disc and those in the buldge, whereas the yellow symbiotic stars (low metallicity, halo systems with a K giant donor) may split into two populations, with $P_{\rm orb} \sim 200$–600 days, and $\geq 900$ days, respectively. The symbiotic systems in the Magellanic Clouds have $P_{\rm orb} \geq 900$ days which is consistent with the longer period tail of the galactic S-type systems. The cool giants masses peak around $6\,{\rm M}_\odot$, and the white dwarf masses – around $0.5\,{\rm M}_\odot$, respectively. More detailed discussion of the orbital parameters is included in Mikołajewska (2003; 2007).

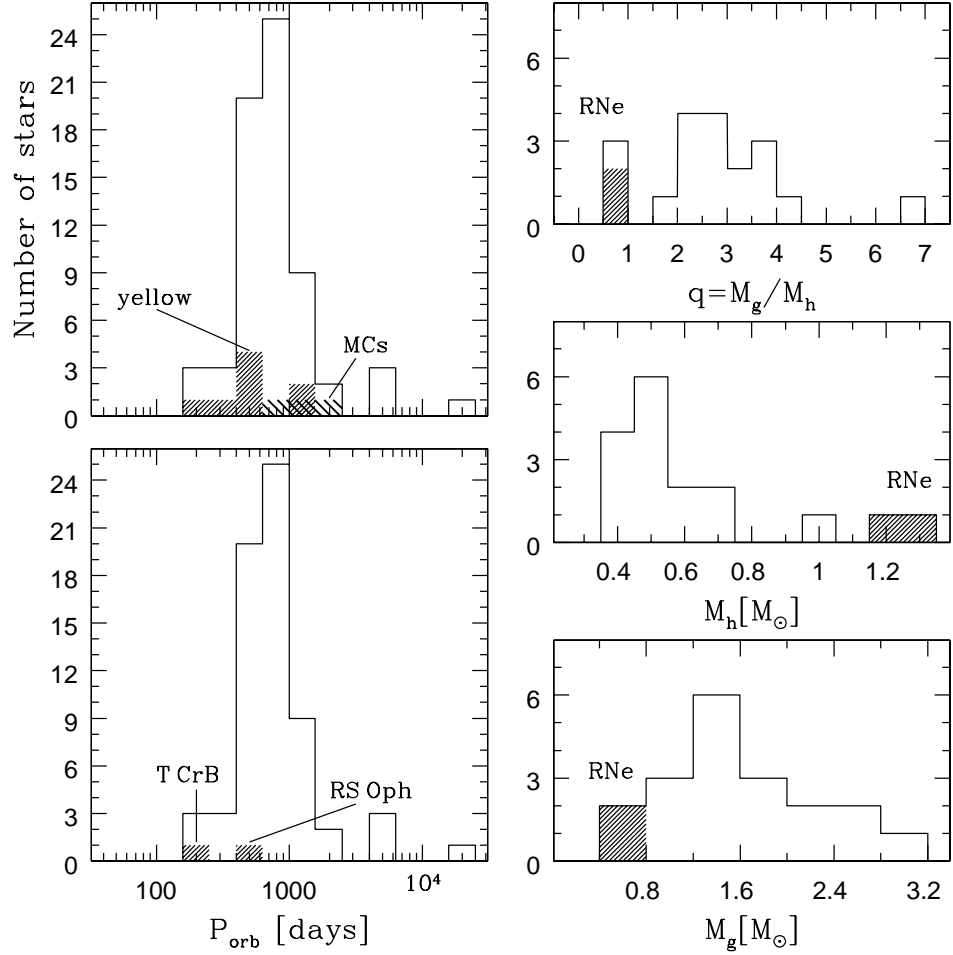

Figure 1. Orbital parameters of symbiotic stars. The shaded regions denote different populations: the MCs, yellow, and symbiotic recurrent novae, respectively.

The orbital parameters are available only for 2 SyRNe: T CrB and RS Oph (e.g. Belczynski & Mikołajewska 1998; Brandi et al. 2008). Both have orbital periods – 227 and 453 days, respectively – consistent with the shorter period tail of galactic S-type systems. There are, however, significant differences between the SyRNe and the other symbiotic stars. In both RS Oph and T CrB, the cool giant is the less massive component, with mass, $M_{\rm g} \sim 0.6$–$0.8\,{\rm M}_\odot$, lower than that of any other symbiotic giant, whereas their white dwarfs are the most massive, with masses, $M_{\rm g} \sim 1.1$–$1.4\,{\rm M}_\odot$, sufficient for them to become Ia supernovae.

## 3. The hot component and its activity

The symbiotic hot components have been recently discussed by Mikołajewska (2003, 2007). According to their activity all symbiotic stars can be split into two subclasses: ordinary or classical symbiotic stars and symbiotic novae.

The vast majority of the symbiotic systems belongs to the first type, including both non-eruptive systems like RW Hya and SY Mus, and systems with Z And-type activity. Their quiescent hot components appear to be relatively hot ($\gtrsim 10^5$ K) and luminous ($\sim 1000\,{\rm L}_\odot$) white dwarfs powered by more or less stable thermonuclear (TNR) burning of the accreted hydrogen. Z And and many other classical symbiotic stars show occasional 1–3 mag optical/UV eruptions on timescales from months to a few years, when the hot component maintains a roughly constant luminosity, whereas its effective temperature varies from $\sim 10^5$ to $\sim 10^4$ K. This activity can be best explained by the presence of an unstable accretion disc around the TNR-shell burning white dwarf (Mikołajewska 2003; Sokoloski et al. 2006). These classical symbiotic white dwarfs cluster around the mass-luminosity relations for stars leaving the AGB with a CO core and those leaving the RGB with a degenerate He core for the first time (see Fig. 5 of Mikołajewska 2003) which suggests that they could still be hot at the onset of mass transfer from the red giant and symbiotic activity.

The symbiotic novae are thermonuclear novae in a symbiotic binary system. There are eight known symbiotic novae, occuring in both S-type (AG Peg, RT Ser, V1329 Cyg and PU Vul) and D-type systems (RR Tel, V1016 Cyg, HM Sge and RX Pup). Their outbursts develop very slowly: the rise to maximum takes months to years, and the decline to the pre-outburst stage can last decades (see Fig. 6 of Mikołajewska 2003). The record-holder among them is AG Peg: its hot component maintained a constant luminosity, $\sim 3000\,{\rm L}_\odot$ for at least 100 years (Kenyon et al. 1993). The evolution of RX Pup was much faster: the constant-luminosity (plateau) phase lasted for only 11 years, and the maximum plateau luminosity, $\sim 15\,000\,L_\odot$, was about 5 times higher than that of AG Peg (Mikołajewska et al. 1999). These differences in outburst evolution reflect different white dwarf masses: higher in RX Pup than that in AG Peg. RX Pup is also a possible recurrent nova (Mikołajewska et al. 1999). The SyRNe are closely related to these symbiotic novae. The main difference between the two groups is the presence of the very massive white dwarfs in the SyRNe which accounts for both recurrence and very short timescales of their nova outbursts.

Between the TNR nova outbursts, the hot components of the SyRNe (including RX Pup) show intrinsic variability resembling the high and low stages of the accretion-powered systems of CH Cyg and V694 Mon (MWC 560) as well as some of the classical (Z And) symbiotic activity, in particular, the activity of all these systems is characterized by very similar timescales (Anupama & Mikoajewska 1999; Mikołajewska 2003, and ref. therein; Gromadzki, Mikołajewska & Lachowicz 2008). Both in the SyRNe and CH Cyg a variable B/A/F-type shell source with $L_{\rm UV/opt} \sim 10\text{--}500\,L_\odot$ appears during the bright (high) state. H I Balmer and He I emission lines are also present but He II and other high-ionization lines are rarely observed. The emission line fluxes require a rather hot source with $T \gtrsim 50\,000$ K and the EUV luminosity, $L_{\rm EUV} \sim L_{\rm UV/opt}$. Similar, double-temperature structure is found in the more luminous, active classical symbiotic stars. In AX Per, AR Pav, FN Sgr, and possibly other systems the shell absorption lines trace the orbit of the hot component and are most likely formed in a geometrically and optically thick accretion disc and gas stream (Quiroga et al. 2002). In RS Oph and T CrB, the hot component brightening is associated with the appearance of rapid photometric variability – flickering, similar to that observed in cataclysmic variables (Anupama & Mikołajewska 1999; Gromadzki, Mikołajewska & Lachowicz 2008). The same correlation between flickering and activity was found in CH Cyg and V694 Mon, and even in the classical symbiotic system Z And during its last series of eruptions (Sokoloski & Bildstein 1999). All these apparent similarities suggest that both the multiple outburst activity of Z And-type symbiotic stars and the high and low states of the SyRNe are due to unstable disc-accretion onto the white dwarf, with the only difference that the white dwarfs in the former burns the accreted hydrogen more or less stably but they do not in the latter.

## 4. The cool giant

The spectral types of the symbiotic giants are now relatively well estimated based on red/near infrared observations. The comparison between symbiotic and single red giants in the solar neighborhood reveals a strong bias towards later spectral types in the former – the spectral type distribution peaks at M5 for S-types, and at M6/M7 for symbiotic Miras, and that the frequency of Mira variables is higher among symbiotic giants (Muerset & Schmid 1999). The symbiotic Miras have also systematically longer pulsation periods than single galactic Miras, and they are surrounded by optically thick dust shells (Whitelock 2003). This predominance of very late, and thus more evolved giants in symbiotic binaries suggests that large radius and high mass loss is essential for triggering symbiotic behaviour in binaries. Indeed symbiotic giants tend to have higher mass loss rates – $\sim 10^{-7}$ and $\sim 10^{-6}\text{--}10^{-5}\,M_\odot\,{\rm yr}^{-1}$ in S- and D-types, respectively – as compared to average red giants and Miras (e.g. Mikołajewska 2003, and references therein). The cool components of the SyRNe follow at least some of these trends. Their spectral types – M2 in RS Oph, M4 in T CrB, and M5–6 in V3890 Sgr and V745 Sco (Anupama & Mikołajewska 1999), fall into the range covered by the S-type systems whereas the short reccurence time for their nova eruptions requires a high accretion rate, $\sim 10^{-7}\,M_\odot\,yr^{-1}$ and consequently a high mass loss rate from the giant.

Although the symbiotic giants are persistently classified as normal luminosity class III giants, their high mass loss rates suggest that they might be more evolved than average field giants with similar spectral types. Recently, Gromadzki et al. (2007) have shown that most S-type symbiotics contain low-amplitude SRb variables instead of normal (nonvariable) giants which can account for their high mass loss rates.

Unfortunately, there is only one symbiotic giant, CH Cyg with a directly measured radius. The IOTA interferometric observations gave $R_g = 300\,R_\odot$ in 1996 (Dyck et al. 1998), and $250\,R_\odot$ (Weigelt et al.; Yudin 2002, private communication) in 2001, respectively. This is a factor of 2 more than an average radius for an M6/M7 III star. One must note, however, that the measured radii for cool giants show large scatter, and the mean standard deviations for the resulting average radii are 50 % (van Belle et al. 1999).

Recently, Zamanov et al. (2007) have measured the projected rotational velocities, $v_{rot}\sin i$, for the giants in 29 S-type symbiotics and demonstrated that these giants are synchronised with the orbital motion. Their result is consistent with theoretical studies of tidal synchronization which predict the synchronization timescales of $\sim 10^3$–$10^4$ yr for typical parameters of S-type symbiotics (e.g. Zamanov et al. 2007, and references therein). At present, there are 48 symbiotic giants with published projected rotational velocities (Kenyon et al. 1991; Mikołajewska & Kenyon 1992; Fekel et al. 2003; Hinkle et al. 2006; Zamanov et al. 2007), and 31 of them also have known orbital periods (Mikołajewska 2003; Gromadzki et al. 2007, in preparation). So, assuming co-rotation with the orbital period for all giants in S-type symbiotics, their radii can be in principle estimated from $v_{rot}\sin i$. Fig. 2a compares these radii with the average radii corresponding to the giant spectral types. Most of systems with known orbital periods are eclipsing binaries, and so have $\sin i \geq 0.94$, and in any case are unlikely to have $\sin i \leq 0.75$ (see discussion in Mikołajewska 2007). Thus we have assumed $\sin i = 1$. In the case of BX Mon and CD-43°14304 – systems with significantly eccentric orbits, we have adopted the pseudo-synchronization periods (Hut 1981) instead of the orbital ones.

Generally the radii derived from $v_{rot}\sin i$ agree with those predicted by spectral type. However, in some systems they are significantly larger, although consistent with the tidal (Roche-lobe) radii. The optical measurements of $v_{rot}\sin i$ give systematically larger values than the near-IR, and there are large differences in some systems with more than one measurement.

The 'rotational' radii of the cool components of both SyRNe deviate significantly from the values predicted by their spectral types. In the case of RS Oph, either the giant rotates faster than the orbital period (Zamanov et al. 2007) or it is filling its tidal lobe. In T CrB, the giant indeed fills its tidal lobe, and its radius resulting from the light curve synthesis is a factor of $\sim 2$ larger than the 'rotational' radius. A similar effect is observed in a few other symbiotics with tidally distorted giants – see below (Sec. 5).

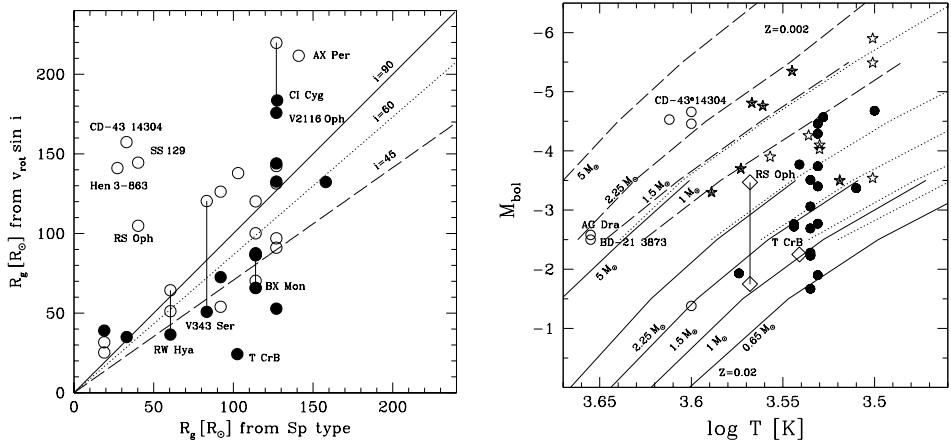

Figure 2. (a) Comparison of the red giant radii derived from $v_{\rm rot} \sin i$ (assuming $\sin i = 1$) with the average radii predicted from their spectral types. Open and filled circles represent the values derived from the optical and near-IR data, respectively. The solid, dashed and dotted lines correspond to $R(v_{\rm rot} \sin i) = R(\rm Sp)$ for $i = 90$, 60 and $45°$, respectively. In the case of T CrB, we have adopted the radius derived from light curve analysis (Belczyński & Mikołajewska 1998) instead of that predicted by the spectral type. (b) Symbiotic giants in the HR diagram. Open and filled circles correspond to the Galactic yellow and remaing systems, respectively. The MC systems are plotted as open (SMC) and filled (LMC) stars. The SyRNe are plotted as open diamonds. The upper and lower position of RS Oph corresponds to the Roche-lobe filling giants, and to the red giant at $d = 1.4$ kpc, respectively (see text). Evolutionary tracks for low-mass AGB stars (dashed and dotted lines for Z=0.002 and Z=0.02, respectively) and RGB stars (solid lines) are also plotted.

The red giant radii combined with the effective temperatures from the spectral type have been used to calculate the luminosity, and to plot the symbiotic giants in an HR diagram (Fig. 2b). In the case of the MC's symbiotics the distances are known, and their luminosities have been estimated from K magnitudes. Both the galactic and the MC symbiotic giants are low-mass, $M_{\rm g} \lesssim 2 - 3 \, \rm M_\odot$, objects, either on the AGB – in the MC systems and yellow Galactic systems with $P_{\rm orb} \gtrsim 900 \, \rm d$, or at the top of RGB or the bottom of the AGB in the remaining systems (see also discussion in Mikołajewska 2007). The position of the cool component T CrB in the HR diagram has been estimated assuming a Roche lobe filling giant, and it is consistent of a low mass, $\lesssim 1 \, \rm M_\odot$, solar metallicity giant. In the case of RS Oph, the upper position corresponds to the Roche lobe filling giant (which is consistent with its rotational velocity) whereas the lower one is consistent with the commonly accepted distance, $d \sim 1.4$ kpc (Rich et al. 2008). In both cases, the location of RS Oph in the HR diagram is consistent with a low mass, $\lesssim 1 \, \rm M_\odot$, giant only if its metallicity is significantly subsolar.

Metal poor giants with [Fe/H]$\lesssim -1$ are present in the yellow symbiotic systems (e.g. Mikołajewska 2007, and references therein), and they all belong to the Galactic halo population. So, is RS Oph another yellow symbiotic star? Scott et al. (1994) found that near-IR spectra of RS Oph indicate very low [C/H] $\lesssim -3$ and $^{12}$C/$^{13}$C=10, and these values are very similar to those estimated for metal-poor ([Fe/H] $\sim -2$) field halo giants (e.g. Keller, Pilachowski & Sneden 2001). However, recent studies of the optical spectra of T CrB and RS Oph have given a nearly solar metallicity (Wallerstein, Harrison & Munari 2006). The only chemical anomaly detected in these two SyRNe is the lithium overabundance (Shabhaz et al. 1999; Wallerstein et al. 2006). Such high Li abundances are common in late-type secondaries in neutron star and back hole binaries (e.g. Martin et al. 1994) but extremely rare in the symbiotic giants. In fact, the Li enhancement is detected only in the symbiotic Mira V407 Cyg, where it can probably be explained as a consequence of hot bottom burning, which occurs in stars with initial masses in the range 4–6 $M_\odot$ (Tatarnikova et al. 2003). The very long pulsation period, $P = 745$ days, supports this interpretation in the case of V407 Cyg as similar Li enhancements have been found in LMC miras with very long periods ($\gtrsim 400$ days), and in one galactic mira with comparable period. A similar explanation is, however, unlikely in the case of the low mass, $\lesssim 1\,M_\odot$, nonpulsating giants in the SyRNe. The fact that the Li enhancement is a common feature of the secondaries in both low-mass X-ray binaries (LMXRB), and the SyRNe, indicates that there must be a process of Li production operating in such binary systems, and independently of the very different nature of their compact companions (a black hole of a neutron star in LMXRB, and a massive white dwarf in the SyRNe).

## 5. Mass transfer mode and accretion

Among the most fundamental questions posed by the symbiotic binaries is the mode of mass transfer – Roche-lobe overflow or stellar wind, and the possibility of accretion disc formation.

3-D hydrodynamical models have shown that in typical symbiotic binaries the cool giant wind is significantly deflected towards the orbital plane by the gravitational pull of its companion (e.g. Mastrodemos & Morris 1998, 1999; Gawryszczak, Mikołajewska & Różyczka 2002, 2003) which would also facilitate the formation of an accretion disc around the companion. Such a picture is supported by observations. In particular, resolved nebulae in D-type symbiotics show a bipolar geometry which is best accounted by intrinsic asphericity of the wind and/or effects associated with the presence of an accretion disc around the hot component. There is also strong spectroscopic evidence for fast, $\gtrsim 1000\,\mathrm{km\,s^{-1}}$, jets and bipolar outflows in active S-type symbiotics (e.g. Tomov 2003) including Z And (Burmeister 2008) most likely produced in an accretion disc environment. So, at least transient accretion discs seem to be commonly formed in these binaries.

However, till recently, most researchers have favored wind accretion over the Roche-lobe overlow. Whereas this standpoint is out of question in the case of D-type systems, the situation in S-type systems, especially those with shorter, $\lesssim 1000^{\rm d}$, orbital periods, is less obvious. In fact, the main arguments against the Roche-lobe overflow have been the red giant radii derived from their projected rotational velocities, and the lack of evidence for ellipsoidal variability in their light curves. The latter argument is not true anymore as there is clear evidence for ellipsoidal light curve variability in many S-type systems which would suggest that the mass-losing giant is filling its Roche lobe or is at least very close to filling its Roche lobe. In particular, tidally distorted giant donors have been detected in all symbiotics with multiple outburst Z And-type activity whenever relevant red or near-IR light curves have been available (Mikołajewska 2007). Moreover, in most of the symbiotics with ellipsoidal variability there is significant discrepancy between the red giant radii derived from $v_{\rm rot} \sin i$ and those indicated by the light curve analysis (see Fig. 4 of Mikołajewska 2007, and discussion therein). Although there is not yet a good solution to this problem, one should remember that the $v_{\rm rot} \sin i$ estimates for these tidally distorted symbiotic giants are biased because the usually adopted model assumes a sherical shape of the giant and a simple limb-darkening law (Orosz & Hauschildt 2000).

The ellipsoidal variations are definitely present in the quiescent optical and near-IR light cuves of T CrB, and the light curve analysis have confirmed that the cool component fills its tidal lobe (Belczynski & Mikołajewska 1998). In the case of RS Oph, the situation is more complicated. While the cool giant radius inferred from the $v_{\rm rot} \sin i$ measurement indicates that it may fill its Roche lobe, the quiescent visual light cuves show very complex orbital behaviour (Gromadzki, Mikołajewska & Lachowicz, 2008). In particular, there is always present a minimum at the time of the spectroscopic conjunction with the red giant in front, and the depth correlated with the system average brightness. There is also a moving bump similar to those found in other active symbiotic systems but no evidence for a secondary minimum expected in the case of the Roche-lobe filling giant. One should note, however, that the moving bump can very easily veil the secondary minimum as it does for instance in AR Pav in which the ellipsoidal changes are very evident in the near-IR bands, whereas the visual light curve shows only very deep primary minimum and the bump (Rutkowski, Mikołajewska & Whitelock 2007). Although it is possible that the cool component of RS Oph does not fill it Roche lobe, it is hard to understand why the evolutionary status and the mass transfer mode of RS Oph differ so much from those in T CrB, especially given that RS Oph is more active and its hot component brigher, and it has a much shorter recurrence time between the TNR eruptions than T CrB. This requires a higher mass transfer/accretion rate in RS Oph than that in T CrB, which is easier to ensure by the Roche-lobe overflow.

## 6. Concluding remarks

The main points of this paper are summarized below.

- The recurrent novae with red giant secondaries are definitely symbiotic stars as they share many physical characteristics with these systems.

- The orbital periods of the SyRNe, RS Oph and T CrB fall into the shorter period tail of galactic S-type systems. These two SyRNe, and V2116 Oph/GX1+4 are the only symbiotics with the cool giant being the less massive component, with mass $\lesssim 1\,{\rm M}_\odot$ in all cases. The white dwarfs in the SyRNe are the most massive, with $M \sim 1.1 - 1.4\,\mathrm{M}_\odot$, sufficient for them to become supernovae Ia.

- Roche-lobe overflow seems to be quite common in S-type symbiotic stars, especially in those with multiple outburst Z And-type activity, and the tidally distorted red giant and Roche-lobe overflow is also present in T CrB and possibly in RS Oph.

- Both the activity of Z And-type symbiotics as well as the high and low states observed in the SyRNe between their TNR nova eruptions are due to unstable disc-accretion onto white dwarf. However the white dwarfs in Z And-type systems burn the accreted hydrogen more or less stably whereas in the SyRNe they do not.

**Acknowledgments.** This research was partly supported by KBNgrant 1P03D 017 27.